\documentclass[preprint,aps,12pt,preprintnumbers,eqsecnum,nofootinbib]{revtex4}
\usepackage{xcolor}
\usepackage{graphicx}
\usepackage{subcaption}
\usepackage{epstopdf} 
\usepackage{braket}
\usepackage{amssymb,amsmath,amsfonts,comment,latexsym,graphicx,epstopdf,float}
\usepackage{color}
\usepackage{fancyvrb}
\usepackage{chngcntr}
\usepackage{etoolbox}
\usepackage{array}
\usepackage{slashed}

\usepackage[colorlinks=true,citecolor=blue,linkcolor=red,filecolor=cyan,urlcolor=magenta]{hyperref}

\newcommand{\be}{\begin{equation}}
\newcommand{\ee}{\end{equation}}
\newcommand{\ba}{\begin{eqnarray}}
\newcommand{\ea}{\end{eqnarray}}

\newcommand{\grts}{\raise.3ex\hbox{$>$\kern-.75em\lower1ex\hbox{$\sim$}}}
\newcommand{\lets}{\raise.3ex\hbox{$<$\kern-.75em\lower1ex\hbox{$\sim$}}}

\newcommand{\dd}{\text{d}}

{\catcode`\|=\active\gdef\Braket#1{\left<\mathcode`\|"8000\let|\bravert 
{#1}\right>}}

\def\bravert{\egroup\,\vrule\,\bgroup}

\usepackage{color}
\usepackage{amssymb,amsmath,amsfonts}
\usepackage{epstopdf} 
\usepackage{braket}

\usepackage[export]{adjustbox} 
\newcommand\BubbleDiagram{ \makebox{\raisebox{-1cm}{\includegraphics{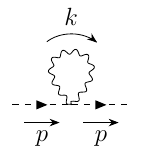}~}} }
\newcommand\RainbowDiagram{ \makebox{\raisebox{-1.1cm}{\includegraphics{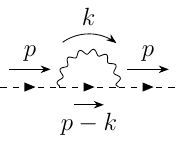}}} }

\begin{document}
%
%
\title{\vspace*{0.5in} 
Asymptotic nonlocality in gauge theories
\vskip 0.1in}
\author{Jens Boos}\email[]{jboos@wm.edu}
\author{Christopher D. Carone}\email[]{cdcaro@wm.edu}
\affiliation{High Energy Theory Group, Department of Physics,
William \& Mary, Williamsburg, VA 23187-8795, USA}
\date{\today}
%
%
\begin{abstract}
Asymptotically nonlocal field theories represent a sequence of higher-derivative theories whose limit point is a ghost-free, infinite-derivative theory.  
Here we extend this framework, developed previously in a theory of real scalar fields, to gauge theories.  We focus primarily on asymptotically nonlocal scalar 
electrodynamics, first identifying equivalent gauge-invariant formulations of the Lagrangian, one 
with higher-derivative terms and the other with auxiliary fields instead.   We then study mass renormalization of the 
complex scalar field in each formulation, showing that an emergent nonlocal scale  ({\em i.e.}, one that does not appear as a fundamental parameter 
in the Lagrangian of the finite-derivative theories) regulates loop integrals as the limiting theory is approached, so that quadratic divergences can be 
hierarchically smaller than the lightest Lee-Wick partner.  We conclude by making preliminary remarks on the generalization of our approach to 
non-Abelian theories, including an asymptotically nonlocal standard model.
\end{abstract}

\maketitle

\section{Introduction}\label{sec:intro}

Quantum field theories involving higher-derivative quadratic terms have been of substantial interest due to the improvement in the short-distance behavior of amplitudes \cite{Boos:2021chb,Stelle:1977ry,Efimov:1967,Krasnikov:1987,Kuzmin:1989,Tomboulis:1997gg,Grinstein:2007mp,Carone:2008iw,Modesto:2011kw,Biswas:2011ar,Ghoshal:2017egr,Buoninfante:2018mre,Boos:2020qgg,Ghoshal:2020lfd}.   Higher-derivative theories can smooth singularities at the origin in nonrelativistic potentials~\cite{Stelle:1977ry}, and can provide solutions to the standard model hierarchy problem, as in the Lee-Wick Standard Model (LWSM)~\cite{Grinstein:2007mp}. Theories involving a small, finite number of higher-derivative quadratic terms, like the LWSM, and ghost-free theories with infinite number of derivatives~\cite{Efimov:1967,Krasnikov:1987,Kuzmin:1989,Tomboulis:1997gg,Modesto:2011kw,Biswas:2011ar,Ghoshal:2017egr,Buoninfante:2018mre,Boos:2020qgg,Ghoshal:2020lfd} have been studied in the literature. In Ref.~\cite{Boos:2021chb}, we proposed a class of theories that interpolates between these possibilities and that may eventually have phenomenological applications.  The purpose of the present work is to explore how the framework of our earlier paper, which focused on a higher-derivative theory of a real scalar field, may be implemented in more realistic quantum field theories.

More explicitly, in Ref.~\cite{Boos:2021chb} we defined a sequence of higher-derivative theories of a real scalar field with a limit point that corresponds 
to an infinite-derivative, ghost-free nonlocal theory.  Such nonlocal theories have quadratic terms involving entire functions of derivatives, so that no new poles 
appear in the two-point function. The sequence of local, higher-derivative theories that approach this limiting theory have a finite but growing number 
of propagator poles, with all but the lightest ({\em i.e.}, the Lee-Wick partners) becoming infinitely heavy as the limit is approached. We called these 
theories ``asymptotically nonlocal'' \cite{Boos:2021chb}.  Asymptotically nonlocal theories are interesting for a number of reasons:  At low energies, these theories exhibit some
features of the nonlocal limiting theory while avoiding the appearance of entire functions of momentum in propagators that lead to complications 
associated with the unitarity of the theory.  We comment on this issue in Sec.~\ref{sec:review}.   Moreover, when loop diagrams in the limiting theory 
are regulated by the nonlocal scale, as in the scalar theory of Ref.~\cite{Boos:2021chb}, one expects that this scale emerges in the finite derivative 
theories that approach it, even though it does not appear as a fundamental parameter in the Lagrangian.  It was shown in the scalar theory of 
Ref.~\cite{Boos:2021chb} that the emergent regulator scale, $M_\text{nl}$, is related to the mass of the lightest Lee-Wick particle, $m_1$, by
\begin{align}
M_\text{nl}^2 \sim {\cal O}\left(\frac{m_1^2}{N}\right) \, ,
\end{align}
where $N$ is the number of propagator poles. This parametric suppression allows one to hold the scale of quadratic divergences fixed while allowing 
the lightest partner particle to be arbitrarily heavy.   

The results of Ref.~\cite{Boos:2021chb} are intriguing, but were only illustrated in a toy model of real scalar fields with a quartic interaction term.  It is natural to 
question whether the qualitative features of the simple scalar theory persist in more realistic ones.  In this paper, we begin addressing this issue by constructing 
and studying asymptotically nonlocal gauge theories.  We focus primarily on an asymptotically nonlocal generalization of scalar quantum electrodynamics (QED).  
Paralleling the approach of Ref.~\cite{Boos:2021chb}, we first present equivalent gauge-invariant formulations of the theory, one with higher-derivative quadratic 
terms and one where these are eliminated in favor of auxiliary fields. To understand mass renormalization, we study the on-shell self-energy for a complex scalar field in 
this theory, in both formulations, and show that the same qualitative behavior found in asymptotically nonlocal $\phi^4$ theory persists in scalar QED.  In particular, we 
demonstrate that the theory is free of a hierarchy problem, with corrections to the squared mass of the complex scalar field set by an emergent nonlocal scale that is hierarchically 
smaller than the lightest Lee-Wick partner as the limiting theory is approached.

We also make some preliminary remarks on non-Abelian theories, including the asymptotically nonlocal generalization of the standard model. Higher-derivative non-Abelian theories have unavoidable derivative interaction terms, so that the resulting theory has logarithmic, but not quadratic, divergences~\cite{Grinstein:2007mp,Carone:2008iw}. We present a plausibility argument for why the dependence on any high cut-off (for example, the Planck scale) should remain logarithmic in such a theory, while the overall scale of radiative corrections to scalar masses should be set by the emergent nonlocal scale, as the asymptotically nonlocal limit is taken. Assuming the emergent nonlocal scale is around a TeV, this could address the hierarchy problem in the standard model while allowing the masses of Lee-Wick resonances to be well beyond the reach of current collider experiments (where they are notably not seen \cite{ParticleDataGroup:2020ssz}). We defer a test of this conjecture, by explicit loop calculations in non-Abelian theories and in the standard model itself, for  future work. We content ourselves here with briefly stating how to define asymptotically nonlocal non-Abelian theories in higher-derivative form, theories that display some of  the distinctive features of the nonlocal limiting theory in the infrared. For example, one would expect deviations from the momentum dependence of tree-level scattering amplitudes in the standard model, which may have experimentally observable consequences.

Our paper is organized as follows: In Sec.~\ref{sec:review}, we review the framework for constructing asymptotically nonlocal theories that was illustrated in a theory of real scalar fields in Ref.~\cite{Boos:2021chb}, and 
summarize the main results of that work. In Sec.~\ref{sec:scalarQED}, we show how the same construction can be generalized to scalar QED, an Abelian gauge theory.  We show how this theory can be written in higher-derivative 
and in Lee-Wick form ({\em i.e.} a form with distinct fields corresponding to each propagator pole, but no higher-derivative terms), and introduce a coupling to a complex scalar field of unit charge.  In Sec.~\ref{sec:self-energy},
we study the radiative corrections to the complex scalar mass, verifying agreement between results computed in the higher-derivative and Lee-Wick descriptions, which are gauge fixed in different ways. We use these results to show that the asymptotically 
nonlocal behavior found in the scalar theory of Ref.~\cite{Boos:2021chb}, {\em i.e.}, a separation between the scale of quadratic divergences (the emergent nonlocal scale) and the mass of the lightest Lee-Wick partner state,  is replicated 
in this gauge theory. In Sec.~\ref{sec:sm}, we briefly discuss the generalization to non-Abelian theories, as well as the complications that arise therein, and we state the full asymptotically nonlocal Lagrangian for the standard model in higher-derivative form, as a point of reference 
for further investigation. In the final section, we summarize our conclusions.

Note that the results of Ref.~\cite{Boos:2021chb} were determined at the one-loop level, but argued to hold at all orders in perturbation theory based on a dimensional argument that we reiterate in Sec.~\ref{sec:review}.  As an 
additional nontrivial consistency check, we provide an appendix in which we show by direct calculation that the conclusions of Ref.~\cite{Boos:2021chb} remain unchanged when two-loop effects are taken into account.

\section{Review of asymptotic nonlocality in a scalar theory} \label{sec:review}

To illustrate our approach, we review the asymptotically nonlocal theory of real scalar fields presented in Ref.~\cite{Boos:2021chb}.
Consider the following Lagrangian of $N$ real scalar fields $\phi_j$, and $N-1$ real scalar fields $\chi_j$,
\begin{equation}
{\cal L}_N = -\frac{1}{2} \, \phi_1 \Box \phi_N - V(\phi_1) - \sum_{j=1}^{N-1} \chi_j \, \left[ \Box \phi_j - (\phi_{j+1}-\phi_j)/a_j^2\right] \,\,\, ,
\label{eq:start}
\end{equation}
where the constants $a_j$ have units of length.   We have set the coefficients of the terms involving $\chi_j$ to one without loss of generality;
this choice may be achieved by rescalings of the $\chi_j$, as these fields do not appear anywhere else in the Lagrangian. Integration over 
the $\chi_j$ leads to functional delta functions in the generating functional for the theory.  This allows one to eliminate $\phi_j$, for 
$j=2 \ldots N$, via the constraints
\begin{equation}
\Box \phi_j -(\phi_{j+1}-\phi_j)/a_j^2=0 \,\,\, ,\,\,\,\,\,\, \mbox{ for }j=1 \dots N-1.
\label{eq:recursive}
\end{equation}
In particular, this implies
\begin{equation}
\phi_N = \left[\prod_{j=1}^{N-1} \left(1+\frac{\ell_j^2 \Box}{N-1} \right) \right] \phi_1 \,\,\, ,
\end{equation}
where $\ell_j^2 \equiv (N-1) \, a_j^2$, allowing one to rewrite Eq.~(\ref{eq:start}) as 
\begin{equation}
\label{eq:hd-lagrangian}
{\cal L}_N = - \frac{1}{2} \phi_1 \Box \left[\prod_{j=1}^{N-1} \left(1+\frac{\ell_j^2 \Box}{N-1} \right) \right] \phi_1 - V(\phi_1)  \,\,\, .
\end{equation}
If one takes the limit in which the $\ell_j$ approach a common, fixed value, $\ell$, while $N$ is taken to infinity, then this Lagrangian approaches the asymptotic form
\begin{equation}
{\cal L}_\infty = -\frac{1}{2} \, \phi_1 \, \Box \, e^{\ell^2 \Box} \, \phi_1 - V(\phi_1) \,\, .
\label{eq:Linf}
\end{equation}
Nonlocal quantum field theories like Eq.~(\ref{eq:Linf}) that involve the exponential of the $\Box$ operator\footnote{
As we noted in Ref.~\cite{Boos:2021chb},  the $\Box$ operator appearing in the derivation above can be replaced by any $\chi_j$-independent differential 
operator ${\cal D}$.} have been studied extensively in the literature; see Ref.~\cite{Buoninfante:2018mre} and many historical references therein. Since the propagator in such theories involves an exponential of a momentum $p$, which diverges in some directions in the complex $p^0$ plane, 
the usual assumption of vanishing contours at infinity that allows for Wick rotation cannot be justified, and there can be a loss of unitarity in simple theories formulated 
in Minkowski space~\cite{Carone:2016eyp}.   One approach is to define such theories in Euclidean space from the start, and only analytically continue amplitudes to Minkowski space after
the loop integrals have been evaluated.   (This was noted in Ref.~\cite{Carone:2016eyp}, but is in fact an assumption that is implicit 
in much of the phenomenological literature on these theories; see the related discussions in Refs.~\cite{Pius:2016jsl,Briscese:2018oyx,Briscese:2021mob,Koshelev:2021orf}.) Asymptotically nonlocal
theories defined in Minkowski space do not present these difficulties and unitarity is maintained via the same prescriptions for handling poles in the complex $p^0$ plane that are employed in other Lee-Wick theories~\cite{Cutkosky:1969fq,Anselmi:2017yux}.  As noted in Ref.~\cite{Boos:2021chb}, the asymptotically nonlocal theory defined by Eq.~(\ref{eq:hd-lagrangian}) has loop integrals that are regulated by an emergent scale $M_{\rm nl}^2 \sim {\cal O}(1/\ell^2)$ 
that is approached for large-but-finite $N$; this result can be anticipated since our higher-derivative $\phi^4$ model is a finite theory and $\ell$ is the only dimensionful scale appearing in the limiting form of the Lagrangian that could serve as a regulator.   It was demonstrated in Ref.~\cite{Boos:2021chb} that the scalar self-energy $M^2(k^2)$
that follows from Eq.~(\ref{eq:hd-lagrangian}) in a massless theory with the choice $V(\phi_1) = \lambda \, \phi_1^4 / 4!$  has the limit
\begin{equation}
\lim\limits_{N\rightarrow\infty} M^2(k^2) = \frac{\lambda}{32\pi^2\ell^2} = \frac{3 \lambda \, M_\text{nl}^2}{128\pi^2} \, ,
\label{eq:nlcutoff}
\end{equation}
when one parametrizes the Lee-Wick partner spectrum of the theory by
\begin{align}
\label{eq:solvable-model-mass-parametrization}
m_j^2 = \frac32 \frac{1}{2 - \frac{j}{N}} N M_\text{nl}^2  \,\,\, ,\,\,\,\,\,\, \mbox{ for }j=1 \dots N-1,
\end{align}
where $m_j \equiv 1/a_j$, and we define $M_\text{nl}^2 \equiv 4/(3 \,\ell^2)$,  so that $(N-1) \,a_j^2 \rightarrow \ell^2$ for $N \rightarrow \infty$, as indicated earlier.
 [In fact, the result in Eq.~(\ref{eq:nlcutoff}) was not sensitive to the form of the parametrization provided that the masses approach a 
common value as they are taken to infinity.]  Equations~(\ref{eq:nlcutoff}) and (\ref{eq:solvable-model-mass-parametrization}) make clear that the nonlocal mass scale can remain fixed at any desired value as one increases $N$, while the massive states become hierarchically heavier. A discussion of mass renormalization at two loops in the $\phi^4$ theory of Ref.~\cite{Boos:2021chb} is included in an 
appendix as further evidence of the robustness of this qualitative result.

Note that the parametrization in Eq.~(\ref{eq:solvable-model-mass-parametrization}) was chosen to assure a spectrum of states that is nondegenerate for any finite $N$. As a consequence, the propagator 
\begin{equation}
D_F(p^2) = \frac{i}{p^2} \, \prod_{j=1}^{N-1} \left(1-\frac{\ell_j^2 \, 
p^2}{N-1}\right)^{-1}\,\,\, .
\label{eq:nondegprop}
\end{equation}
that follows from Eq.~(\ref{eq:hd-lagrangian}) can be decomposed via partial fractions as a sum over simple poles with finite residues that alternate in sign.  This is precisely the
expectation in theories with higher-derivative quadratic terms~\cite{Pais:1950za}, and has been illustrated previously in generalizations of the LWSM that involve more than one Lee-Wick partner state~\cite{Carone:2008iw}.

We do not repeat here the one-loop calculation in $\phi^4$ theory that establishes Eq.~(\ref{eq:nlcutoff}) with $M_{\rm nl}^2 \sim {\cal O} (m_1^2/N)$, where $m_1$
is the mass of the lightest Lee-Wick partner.  We refer the reader to Ref.~\cite{Boos:2021chb} for details.  Nevertheless, a similar (though more nontrivial) study of the 
one-loop contributions to the 
self-energy in an Abelian gauge theory is discussed in depth in Sec.~\ref{sec:self-energy}, with calculations presented in both the higher-derivative and 
Lee-Wick forms of the theory.   We will see explicitly that an emergent cutoff is again obtained that is hierarchically lighter than the Lee-Wick partners.

\section{Asymptotically nonlocal Abelian gauge theories} \label{sec:scalarQED}

Motivated by the results of Ref.~\cite{Boos:2021chb}, which we have summarized in the previous section, we now present a generalization to an Abelian gauge theory. As in Sec.~\ref{sec:review}, we first discuss the auxiliary Lagrangian, which involves a generalization of the $\chi_j$ fields,
 and then consider the higher-derivative theory that results from integrating out the associated constraints.  Alternatively, we show that it is possible to recast the auxiliary theory into Lee-Wick form by using field redefinitions and integrating out the remaining nondynamical fields.  Since the field redefinitions are identical to those encountered in the scalar model of Ref.~\cite{Boos:2021chb}, we may use these results to aid us in extracting the physical content of the theory we consider here. Finally, in the limit of infinitely many auxiliary fields, we show that the theory becomes asymptotically nonlocal and is also free of a hierarchy problem.  Extending this theory to non-Abelian theories and the standard model is discussed in Sec.~\ref{sec:sm}.

Let us first focus on the pure gauge field part of the Lagrangian; the gauge-fixing will be discussed shortly, and the coupling to matter fields will be covered at the end of this section. The Lagrangian contains $N$ vector fields $\hat{A}_\mu^j$ as well as $N-1$ auxiliary vector fields $\chi_\mu^j$:\footnote{Here and in what follows we will either use $\hat{A}^j_\mu$ or $\hat{A}_j^\mu$, that is, Lorentz indices are raised and lowered as usual with the Minkowski metric, but we will place the index $j$ that numbers the vectors wherever it is most convenient in order to improve the readability of our formulae.}
\begin{align}
\label{eq:aux-lagrangian}
\mathcal{L}^\text{gauge}_N = \frac12 \hat{A}{}^1_\mu \mathcal{O}{}^\mu_\nu \hat{A}{}^\nu_N + \sum\limits_{j=1}^{N-1} \chi_\mu^j \left[ \mathcal{O}{}^\mu_\nu \hat{A}{}^\nu_j - \frac{1}{a_j^2}\left(\hat{A}{}^\mu_{j+1} - \hat{A}{}^\mu_j\right) \right] \, .
\end{align}
In the above, $a_j$ are positive constants and the operator $\mathcal{O}^\mu_\nu$ is given by
\begin{align}
\mathcal{O}^\mu_\nu \equiv \delta{}^\mu_\nu \Box - \partial{}^\mu \partial{}_\nu \, .
\end{align}
The Lagrangian is invariant under the U(1) gauge transformation
\begin{align}
\label{eq:abelian-gauge-trafo}
{\hat{A}}_\mu^j \rightarrow {\hat{A}}_\mu^j + \partial_\mu \lambda \, , \quad
\chi^j_\mu \rightarrow \chi^j_\mu \, .
\end{align}
where $\lambda \equiv \lambda(x^\nu)$ is an arbitrary gauge function. The operator $\mathcal{O}^\mu_\nu$ has the following properties:
\begin{itemize}
\item[(i)] $\mathcal{O}^\mu_\nu {\hat{A}}{}^j_\mu$ is gauge invariant under \eqref{eq:abelian-gauge-trafo}, and ${\hat{A}}{}^j_\mu \mathcal{O}^\mu_\nu$ times any product of the fields is gauge invariant up to a surface term.\\[-1.5\baselineskip]
\item[(ii)] $\mathcal{O}^\mu_\nu$ gives rise to a Maxwell-like kinetic term, $\frac12 {\hat{A}}{}^j_\mu \mathcal{O}^\mu_\nu {\hat{A}}{}_j^\nu = -\frac14 F{}_{\mu\nu}^2$, up to surface terms, where $F_{\mu\nu} = \partial_\mu {\hat{A}}^j_\nu - \partial_\nu {\hat{A}}^j_\mu$.\\[-1.5\baselineskip]
\item[(iii)] The operator satisfies $(\mathcal{O}^\mu_\nu)^n \equiv \mathcal{O}^\mu_{\nu_1} \mathcal{O}^{\nu_1}_{\nu_2} \cdots \mathcal{O}^{\nu_{n-1}}_\nu = \mathcal{O}^\mu_\nu \Box^{n-1}$, which will be useful later.
\end{itemize}
Notice that (i) above requires that the $\chi^j_\mu$ transform as singlets to maintain the invariance of the first term in the sum, even if each $\hat{A}^j_\mu$ were able to shift under different U(1) gauge transformations; however, invariance of the second term in the sum then requires that all the $\hat{A}^j_\mu$ shift by a common gauge function, as expected for a theory invariant under a single U(1) gauge group. It then follows that the off-diagonal kinetic term involving $\hat{A}^1_\mu$ and $\hat{A}^N_\mu$ is gauge invariant as well, establishing the U(1) gauge invariance of the Lagrangian as a whole.

While the Lagrangian in Eq.~\eqref{eq:aux-lagrangian} appears non-generic, this is somewhat misleading since this expression should not be thought of as the fundamental description of the theory.  As we will see in the next section, Eq.~\eqref{eq:aux-lagrangian} is equivalent to a higher-derivative theory in which the quadratic terms are a generic polynomial in the $\Box$ operator, up to the physical constraint that its zeros are real and positive, as parametrized by the $1/a_j^2$. These will correspond to particle squared masses.  To extract the physical interpretation of Eq.~\eqref{eq:aux-lagrangian}, it is instructive to follow two paths: the interpretation of the theory as (i) a higher-derivative modification of Maxwell theory, and (ii) a Lee-Wick theory, with additional particles that are partners to the massless photon. As will become apparent, each perspective can be of value in different circumstances.

\subsection{Higher-derivative picture}
Since the $\chi_\mu^j$ fields only appear linearly in the Lagrangian \eqref{eq:aux-lagrangian}, the functional integral can be performed exactly, giving rise to the $N-1$ constraints
\begin{align}
\hat{A}_{j+1}^\mu = \left( \delta{}^\mu_\nu + a_j^2 \mathcal{O}^\mu_\nu \right) \hat{A}_j^\nu \, , \quad \text{for~} j=1,\dots,N-1 \, .
\end{align}
Inserting this back into the original Lagrangian one can employ the property (iii) to recast
\begin{align}
\mathcal{O}^\mu_\nu \prod\limits_{j=1}^{N-1}\left( \delta{}^\nu_\rho + a_j^2 \mathcal{O}^\nu_\rho \right) = \prod\limits_{j=1}^{N-1}\left( 1 + a_j^2 \Box \right) \mathcal{O}^\mu_\rho \, ,
\end{align}
so that
\begin{align}
\label{eq:gauge-lagrangian}
\mathcal{L}^\text{gauge}_N = \frac12 \hat{A}{}^1_\mu \mathcal{O}{}^\mu_\nu \prod\limits_{j=1}^{N-1}(1 + a_j^2\Box) \hat{A}{}^\nu_1 \, .
\end{align}
This Lagrangian represents a higher-derivative modification of Maxwell theory for the field $\hat{A}_\mu^1$, since, up to surface terms, it corresponds to
\begin{align}
\mathcal{L}^\text{gauge}_N = -\frac14 F_{\mu\nu} \prod\limits_{j=1}^{N-1}(1 + a_j^2\Box) F{}^{\mu\nu} \, , \quad F_{\mu\nu} \equiv \partial_\mu \hat{A}_\nu^1 - \partial_\nu \hat{A}^1_\mu \, .
\end{align}
In order to gauge-fix this Abelian theory we add a standard gauge-fixing Lagrangian
\begin{align}
\mathcal{L}_\text{gf} = -\frac{1}{2\xi} (\partial_\mu \hat{A}^\mu_1)^2 \, .
\label{eq:gfa1}
\end{align}
Then the propagator takes the form
\begin{align}
\label{eq:hd-prop}
\hat{D}^\mu_\nu(p^2) = \frac{-i}{p^2f(p^2)} \left\{ \delta{}^\mu_\nu -\left[1-\xi f(p^2)\right] \frac{p{}^\mu p{}_\nu}{p^2} \right\} \, , \quad f(p^2) \equiv \prod\limits_{j=1}^{N-1}(1-a_j^2p^2) \, .
\end{align}
We decorated the propagator with a hat to indicate that it is the propagator of the higher-derivative theory. Next, it is useful to perform the partial fraction decomposition
\begin{align}
\frac{1}{f(p^2)} = \sum\limits_{j=1}^{N-1} \frac{b_j}{p^2-m_j^2} \, ,
\quad b_j \equiv -m_j^2 \prod\limits_{\substack{k=1\\k\not=j}}^{N-1} \frac{m_k^2}{m_k^2-m_j^2} \, , \quad m_j^2 = \frac{1}{a_j^2} \, .
\end{align}
The coefficients $b_j$ satisfy the following useful relations:
\begin{align}
\sum\limits_{j=1}^{N-1} b_j = 0 \, , \qquad
\sum\limits_{j=1}^{N-1} b_j m_j^{2n} = 0 \quad \text{for~} n=1,\dots,N-2 \, .
\end{align}
It is also convenient to define the quantities 
\begin{equation} \label{eq:cj}
c_j \equiv b_j/m_j^2  \,\, .
\end{equation}
With $c_0=1$, the $c_j$ are the residues of the poles in the partial fraction decomposition of $[p^2f(p^2)]^{-1}$. Defining $m_0 \equiv 0$ they inherit the properties
\begin{align}
\label{eq:cancellation}
\sum\limits_{j=0}^{N-1} c_j = 0 \, , \qquad
\sum\limits_{j=0}^{N-1} c_j m_j^{2n} = 0 \quad \text{for~} n=1,\dots,N-2 \, ,
\end{align}
where the summation is now carried out from $0$ to $N-1$. These technical relations are of central importance for many subsequent conclusions of this paper, which is why we display them here.\footnote{Let us note in passing that these expressions are vaguely reminiscent of Pauli's sum rules from 1951 \cite{Pauli:1951}.}

\subsection{Asymptotic nonlocality}
Going back to the non gauge-fixed Lagrangian, let us introduce a new quantity
\begin{align}
\ell_j^2 = (N-1) \, a_j^2 \, .
\end{align}
Then the $N-1$ constraint equations take the form
\begin{align}
\hat{A}{}^\mu_N = \prod\limits_{j=1}^{N-1} \left(\delta{}^\mu_\nu + \frac{\ell_j^2 \mathcal{O}{}^\mu_\nu}{N-1} \right) \hat{A}{}^\nu_1 \, .
\end{align}
In the limiting case of $N\rightarrow \infty$ one finds (assuming that $\ell_j \rightarrow \ell$) that Eq.~\eqref{eq:gauge-lagrangian} becomes
\begin{align}
\mathcal{L}^\text{gauge}_\infty = \frac12 \hat{A}{}^1_\mu \, e^{\ell^2 \Box} \, \mathcal{O}^\mu_\nu \hat{A}{}^\nu_1 \, .
\end{align}
This is, up to surface terms, the same Lagrangian as the nonlocal Maxwell Lagrangian $-\frac14 F{}_{\mu\nu} e^{\ell^2 \Box}F{}^{\mu\nu}$ that has been studied elsewhere \cite{Buoninfante:2018stt,Boos:2020twu} (see also historical references therein). We have discussed in Sec.~\ref{sec:review} why the large-but-finite-$N$ limit may be preferable to the theory defined at the limit point where $N \rightarrow \infty$ and $(N-1) \, a_j^2 \rightarrow \ell^2$, for all $j$. As we shall show explicitly in Sec.~\ref{sec:self-energy},  quadratic divergences in an Abelian gauge theory with complex scalar fields are regulated by the would-be nonlocal scale $\ell$, which is hierarchically separated from the mass scales $1/a_j^2$ when one approaches this limit at finite $N$.

\subsection{Lee-Wick picture}
Instead of integrating out the $N-1$ auxiliary fields $\chi_\mu^j$ directly, it is also possible to perform a field redefinition and then integrate out the nondynamical fields that remain in this new basis, which gives rise to a Lee-Wick-type theory. Starting with the original Lagrangian \eqref{eq:aux-lagrangian} we define the collection of all fields
\begin{align}
\underline{\hat{A}}_\mu = (\hat{A}{}^1_\mu, \chi{}^1_\mu, \dots, \hat{A}{}^{N-1}_\mu, \chi{}^{N-1}_\mu, \hat{A}{}^N_\mu )
\end{align}
such that the Lagrangian takes the form
\begin{align}
\mathcal{L}^\text{gauge}_N = \frac12 \underline{\hat{A}}_\mu^T \left( K \mathcal{O}{}^\mu_\nu + M \delta{}^\mu_\nu \right) \underline{\hat{A}}^\nu \, ,
\end{align}
where $K$ and $M$ are $(2N-1)\times(2N-1)$ kinetic and mass matrices, and a superscript ``$T$'' denotes transposition. This system can be diagonalized via a field redefinition to a new basis $\underline{\widetilde{A}}_\mu = (A_\mu, \widetilde{A}^1_\mu, \dots, \widetilde{A}^{N-1}_\mu, \widetilde{\chi}^1_\mu, \dots, \widetilde{\chi}^{N-1}_\mu )$ via
\begin{align}
\label{eq:lee-wick-trafo}
\underline{\hat{A}}_\mu = S_N \underline{\widetilde{A}}_\mu \, ,
\end{align}
where $S_N$ is an invertible $(2N-1)\times(2N-1)$ matrix. The matrices $K$ and $M$ are identical to those discussed in our previous paper~\cite{Boos:2021chb}, where explicit forms were presented for $N=2$ and $N=3$. In general, the resulting matrices $K_0 = S_N^T K S_N$ and $M_0 = S_N^T M S_N$ are block-diagonal,
\begin{align} \label{eq:blocks}
\hspace{-5pt}
K_0 = \begin{pmatrix}
1 &        &        &            & 0      \\
  & (-1)^1 &        &            & 0      \\
  &        & \ddots &            & \vdots \\
  &        &        & (-1)^{N-1} & 0      \\
0 &  0     & \hdots & 0          & X
\end{pmatrix} \, , \quad
M_0 = \begin{pmatrix}
0 &              &        &                      & 0      \\
  & (-1)^1 m_1^2 &        &                      & 0      \\
  &              & \ddots &                      & \vdots \\
  &              &        & (-1)^{N-1} m_{N-1}^2 & 0      \\
0 &  0           & \hdots & 0                    & Y
\end{pmatrix} \, ,
\end{align}
where $X$ and $Y$ are $(N-1)\times(N-1)$ blocks that cannot be simultaneously diagonalized and typically depend on arbitrary parameters that enter the field redefinition matrix $S_N$.  This suggests that the corresponding fields 
$\widetilde{\chi}_\mu^j$ with $j=1,\dots,N-1$ are unphysical. By checking concrete expressions for $K_0$ and $M_0$ for various $N$ one can show that it is possible to successively integrate out these auxiliary fields with no effect on the remaining fields $A_\mu$ and $\widetilde{A}_\mu^j$.   For example, in the case of $N=3$ discussed in Ref.~\cite{Boos:2021chb}, a vanishing eigenvalue in $X$, allows one to perform the functional integral over the corresponding $\widetilde{\chi}$ field, leading to a functional constraint that forces the vanishing of the remaining $\widetilde{\chi}$ field.  This pattern must persist for arbitrary $N$ since the physical blocks of $K_0$ and $M_0$ ({\em i.e.}, excluding $X$ and $Y$) are in exact correspondence with the residues and poles of the propagator of the higher-derivative form of the theory. Henceforth, we restrict ourselves to this $N \times N$ subspace which corresponds to a Lee-Wick theory of one massless photon and $N-1$ massive vector partner particles of mass $m_j$. 

In order to develop perturbation theory, we insert a usual gauge-fixing term for the massless photon,
\begin{align}
\mathcal{L}_\text{gf} = -\frac{1}{2\xi} (\partial_\mu A^\mu)^2 \, .
\label{eq:gfa}
\end{align}
As noted in Ref.~\cite{Grinstein:2007mp}, one should obtain the same physical results whether working in the higher-derivative theory, with the field
$\hat{A}_1^\mu$ gauge fixed as in Eq.~(\ref{eq:gfa1}), or in the Lee-Wick form of the theory, with the field $A^\mu$ gauge fixed 
as in Eq.~(\ref{eq:gfa}).  We will see this in our subsequent calculations. The propagator for $A_\mu$ takes the form
\begin{align}
D^\mu_\nu(p^2) = \frac{-i}{p^2}\left[ \delta{}^\mu_\nu - (1-\xi) \frac{p{}^\mu p{}_\nu}{p^2} \right] \, ,
\end{align}
and for the massive vectors $\widetilde{A}_\mu^j$ with $j=1,\dots,N-1$ the propagator can be read off directly,
\begin{align}
\widetilde{D}^\mu_\nu(p^2) = (-1)^j \frac{-i}{p^2-m_j^2}\left( \delta{}^\mu_\nu - \frac{p{}^\mu p{}_\nu}{m_j^2} \right) \, .
\end{align}

\subsection{Coupling to matter} \label{sec:scalarQED:matter}
In order to study the issue of quadratic divergences, we couple the gauge sector to a complex scalar field $\phi$ of unit charge
\begin{align}
\mathcal{L}_\text{matter} = (D_\mu \phi)^\ast(D^\mu \phi) - m_\phi^2\phi^\ast\phi - V(\phi^\ast\phi) \, ,
\label{eq:sqedL}
\end{align}
where the covariant derivative is defined as
\begin{align}
D_\mu \phi \equiv \left( \partial_\mu - i g \hat{A}_\mu^1 \right) \phi \, .
\label{eq:covderiv}
\end{align}
This is unique in the higher-derivative theory. If one started instead with the theory defined in terms of the auxiliary fields, one could imagine 
constructing alternative covariant derivatives in which $\hat{A}_\mu^1$ is replaced with any of the $\hat{A}_\mu^j$; however this corresponds to 
including additional derivative couplings of the gauge field to the matter fields in the higher-derivative description, which is arguably a less minimal 
choice.    In the higher-derivative picture, the coupling to photons given by Eqs.~(\ref{eq:sqedL}) and (\ref{eq:covderiv}) is identical to standard scalar 
QED and the Feynman rules are the same, aside from the differing form of the photon propagator. In the Lee-Wick picture one can show that the first 
row of the field redefinition matrix $S_N$ is given by\footnote{The overall sign of each column of $S_N$ may be changed without altering the diagonal
entries of the matrices in Eq.~(\ref{eq:blocks}).  Our conventions here differ from Ref.~\cite{Boos:2021chb} in that we take the $(S_N)_{0j}>0$, for $j=1,\ldots,N-1$.}
\begin{align} \label{eq:sns}
\begin{split}
\left( S_N \right)_{00} &= 1 \, , \\
\left( S_N \right)_{0j} &= \sqrt{(-1)^j \, c_j} > 0 \, \quad \text{for~} j = 1,\dots,N-1 \, , \\
\left( S_N \right)_{0j} &= 0 \, \hspace{80pt} \text{for~} j = N,\dots,2N-2 \, ,
\end{split}
\end{align}
which implies that the vector field $\hat{A}^1_\mu$ in the original auxiliary theory is related to the massless photon $A_\mu$ and its Lee-Wick partners $\widetilde{A}^j_\mu$ in the Lee-Wick theory as follows:
\begin{align}
\hat{A}^1_\mu = A_\mu + \sum\limits_{j=1}^{N-1} \sqrt{(-1)^j \, c_j} \, \widetilde{A}^j_\mu \, .
\end{align}
Hence the coupling to matter via Eq.~\eqref{eq:covderiv} remains unaffected in the massless gauge sector, whereas the coupling to the Lee-Wick partner vectors $\widetilde{A}_\mu^j$
includes an additional factor of $\sqrt{(-1)^j\,c_j}>0$.

\section{Scalar self-energy} \label{sec:self-energy}

The asymptotically nonlocal $\phi^4$ model discussed in Sec.~\ref{sec:review} provided for the hierarchical separation of the Lee-Wick scale and an emergent nonlocal regulator scale in the decoupling limit, when $N$ becomes large and the ratio $m_j^2/(N-1)$ remains constant. Here we show that the same happens in the Abelian gauge theory of Sec.~\ref{sec:scalarQED}, by considering the one-loop self-energy for the complex scalar field introduced in Sec.~\ref{sec:scalarQED:matter}. We will compute the on-shell self-energy $M^2(m_\phi^2)$ in both the higher-derivative and Lee-Wick forms of the theory to understand how it is regulated. Before delving into the detailed computations, however, let us briefly anticipate the final result:\footnote{When we write logarithms with dimensionful arguments, we can always divide these arguments by an arbitrary dimensionful scale---for example, $m^2_\phi$ in the first logarithm of Eq.(\ref{eq:anticipate})---without changing our results.  This is a consequence of Eq.~(\ref{eq:cancellation}).}
\begin{align} \label{eq:anticipate}
\begin{split}
M^2(p^2=m_\phi^2) &= \BubbleDiagram + \quad \RainbowDiagram \\
&= \frac{g^2}{(4\pi)^2} \sum\limits_{j=1}^{N-1} c_j \left\{ \frac12 \frac{m_j^4}{m_\phi^2} \log m_j^2 + \frac12 \frac{m_j \mu_j^3}{m_\phi^2} \log\left[ \frac{(m_j-\mu_j)^2}{4m_\phi^2} \right] \right\} \, , \\
c_j &\equiv (-1) \prod\limits_{\substack{k=1\\k\not=j}}^{N-1} \frac{m_k^2}{m_k^2-m_j^2} \, , \quad \mu_j \equiv \sqrt{m_j^2-4 \,m_\phi^2} \, .
\end{split}
\end{align}
In the limiting case of vanishing scalar mass, $m_\phi\rightarrow 0$, one obtains the finite result
\begin{align}
M^2(p^2=0) = \frac{3g^2}{(4\pi)^2} \sum\limits_{j=1}^{N-1} c_j m_j^2 \log m_j^2 \, .
\end{align}
The scalar self-energy is manifestly finite, and the scale of the corrections is set by the would-be nonlocal scale $M_\text{nl}$, as we will show below. Since this scale is hierarchically lower than the Lee-Wick masses, this scalar QED model has no hierarchy problem as the Lee-Wick partners are taken heavy. We will discuss this point in more detail below.

\subsection{Higher-derivative computation}
As discussed in Sec.~\ref{sec:scalarQED:matter}, the Feynman rules for the higher-derivative theory are identical to those of scalar QED, with the exception that the gauge propagator takes the form
\begin{align}
\hat{D}^\mu_\nu(p^2) &= -\frac{i}{p^2f(p^2)} \left( \delta{}^\mu_\nu - \frac{p{}^\mu p{}_\nu}{p^2} \right) - i\xi \frac{p{}^\mu p{}_\nu}{p^4} \, , \quad f(p^2) \equiv \prod\limits_{j=1}^{N-1}(1-a_j^2p^2) \, .
\end{align}
Note that the gauge-dependent part proportional to $\xi$ is independent of the higher-derivative modification $f(p^2)$, and hence the question of gauge independence at one loop coincides with that of standard scalar QED. On-shell, the scalar self-energy gives the shift in the physical pole mass and is a manifestly gauge-independent quantity; we will verify this explicitly in the calculations below.

The scalar self-energy is a sum of two diagrams, and the first diagram can be written as
\begin{align}
-i\,{}^{(1)}\!M^2 &\equiv \BubbleDiagram \nonumber \\
&=g^2(d-1)\int\frac{\dd^4 k}{(2\pi)^d}\frac{1}{k^2 f(k^2)} + g^2 \xi \int\frac{\dd^4 k}{(2\pi)^d} \frac{1}{k^2} \\
&= g^2 (d-1) \sum\limits_{j=0}^{N-1} c_j \int \frac{\dd^d k}{(2\pi)^d} \frac{1}{k^2-m_j^2} + g^2\xi\int\frac{\dd^d k}{(2\pi)^d} \frac{1}{k^2} \\
&= -i \Big( \sum\limits_{j=0}^{N-1} {}^{(1)}\!M^2_j + \, M^2_\xi \Big)\, , \\
-i\,M^2_\xi &= g^2\xi\int\frac{\dd^d k}{(2\pi)^d} \frac{1}{k^2} \, ,
\end{align}
where $m_0 \equiv 0$ and $c_0 \equiv 1$ for notational brevity. The gauge-dependent term $M^2_\xi$ formally diverges, but we will see that it is cancelled by the gauge-dependent contribution from the second diagram when the latter is evaluated on-shell, the case of interest. The remaining gauge-independent part is easily evaluated using dimensional regularization in Euclidean space ($\epsilon \equiv 4-d$, $p^0 = ip_E^0$, $p^2=-p_E^2$):
\begin{align}
-i\, \sum\limits_{j=0}^{N-1} \, {}^{(1)}\!M^2_j &= -ig^2 (d-1) \sum\limits_{j=0}^{N-1} c_j \int \frac{\dd^d k_E}{(2\pi)^d} \frac{1}{k_E^2+m_j^2} \\
&= -ig^2 (d-1) \sum\limits_{j=0}^{N-1} c_j \left( -\frac{m_j^2}{8\pi^2} \times \frac{1}{\epsilon} + \text{finite} \right) = -\frac{3ig^2}{(4\pi)^2} \sum\limits_{j=1}^{N-1} c_j m_j^2 \log m_j^2 \, ,
\end{align}
where the $1/\epsilon$-contributions add up to zero due to the cancellation rules \eqref{eq:cancellation}, provided $N \ge 3$. The finite part also picks up contributions proportional to sums over $c_j m_j^2$ which also add up to zero under the summation thanks to \eqref{eq:cancellation}; the argument of the logarithm can be normalized to an arbitrary dimensionful constant for that reason.

The second diagram, evaluated on-shell, is
\begin{align}
-i \, {}^{(2)} \!M^2(p^2=m_\phi^2) &\equiv \RainbowDiagram \nonumber \\
&= -4g^2\int\frac{\dd^d k}{(2\pi)^d} \frac{1}{k^2f(k^2)} \frac{1}{k^2-2p\cdot k}\left[ m_\phi^2 - \frac{(p\cdot k)^2}{k^2} \right] + i\, M^2_\xi \\
&= -4g^2\int\frac{\dd^d k}{(2\pi)^d} \sum\limits_{j=0}^{N-1} \frac{c_j}{k^2-m_j^2} \frac{1}{k^2-2p\cdot k}\left[ m_\phi^2 - \frac{(p\cdot k)^2}{k^2} \right] + i\, M^2_\xi \\
&\equiv -i \, \sum\limits_{j=0}^{N-1} {}^{(2)}\!M^2_j + i \, M^2_\xi \, ,
\end{align}
that is, the gauge-dependent part is identical to the gauge-dependent part of the first diagram, up to a sign, such that they cancel precisely.  Moving to Euclidean space and introducing Feynman parameters one finds
\begin{align}
-i{}^{(2)}\!M^2_j = &-4ig^2m_\phi^2c_j \int_0^1 \dd x \int\frac{\dd^d\ell_E}{(2\pi)^d} \frac{1}{(\ell_E^2 + \Delta^{(a)}_j)^2} \\
&+8ig^2m_\phi^2c_j \int_0^1 \dd x \int_0^{1-x} \dd y  \int\frac{\dd^d\ell_E}{(2\pi)^d} \frac{1}{(\ell_E^2 + \Delta^{(b)}_j)^3} \left( \frac{\ell_E^2}{d} - y^2m_\phi^2 \right)  \, , \nonumber
\end{align}
where we defined the quantities
\begin{align}
\Delta^{(a)}_j &= x m_j^2 + (1-x)^2 m_\phi^2 \, , \quad \Delta^{(b)}_j = x m_j^2 + y^2 m_\phi^2 \, .
\end{align}
Using dimensional regularization we can extract the diverging parts of the Euclidean loop integrals,
\begin{align}
\begin{split}
-i{}^{(2)}\!M^2_j = &-4ig^2m_\phi^2c_j \int_0^1 \dd x \left( \frac{1}{8\pi^2} \times \frac{1}{\epsilon} + \text{finite} \right) \\
&+8ig^2m_\phi^2c_j \int_0^1 \dd x \int_0^{1-x} \dd y  \left[ \frac{1}{32\pi^2} \times \frac{1}{\epsilon} + \text{finite} \right] \, .
\end{split}
\end{align}
The $1/\epsilon$ poles vanish due to the sum rules in Eq.~\eqref{eq:cancellation}.  The finite parts are given by
\begin{align}
\begin{split}
-i{}^{(2)}\!M^2 = \frac{ig^2m_\phi^2}{(4\pi)^2} \sum_{j=0}^{N-1} c_j \Bigg[ &\hspace{13pt} \int_0^1 \dd x \, \left(4\log\Delta^{(a)}_j \right)\\
&- \int_0^1 \dd x \int_0^{1-x} \dd y \left( 2\log\Delta^{(b)}_j + \frac{4y^2m_\phi^2}{\Delta^{(b)}_j} \right) \Bigg] \, .
\end{split}
\end{align}
The final result for the on-shell scalar self-energy is then
\begin{align}
\begin{split}
\label{eq:msquared-hd}
M^2(p^2=m_\phi^2) &= \sum\limits_{j=0}^{N-1} \left( {}^{(1)}M^2_j + M^2_\xi + {}^{(2)}M^2_j - M^2_\xi \right) \\
&= \frac{g^2}{(4\pi)^2} \sum\limits_{j=1}^{N-1} c_j \left\{ \frac12 \frac{m_j^4}{m_\phi^2} \log m_j^2 + \frac12 \frac{m_j \mu_j^3}{m_\phi^2} \log\left[ \frac{(m_j-\mu_j)^2}{4m_\phi^2} \right] \right\} \, , \\
c_j &\equiv (-1) \prod\limits_{\substack{k=1\\k\not=j}}^{N-1} \frac{m_k^2}{m_k^2-m_j^2} \, , \quad \mu_j \equiv \sqrt{m_j^2-4\, m_\phi^2} \, .
\end{split}
\end{align}

\subsection{Lee-Wick computation}

The Feynman rules of scalar QED in the Lee-Wick picture were discussed in Sec.~\ref{sec:scalarQED:matter}; in a nutshell, the photon-scalar vertex remains unchanged, whereas the vertex of the scalar and a massive Lee-Wick partner field $\widetilde{A}^j_\mu$ comes with an additional factor of $\sqrt{(-1)^j c_j}$. Moreover, in this case it is useful to fix the gauge to $\xi=1$ and directly compute the sum of the bubble and rainbow diagram. Here, the external momentum is labelled as $p$, and $d$ denotes the number of spacetime dimensions:
\begin{align}
\begin{split}
-i \,M^2(p^2) &= g^2\int\frac{\dd^d k}{(2\pi)^d} \Bigg\{ \frac{d}{k^2} - \frac{(2p-k)^2}{k^2}\frac{1}{(p-k)^2-m_\phi^2} \\
&+ \sum\limits_{j=1}^{N-1} \frac{c_j}{k^2-m_j^2} \left[ d - \frac{k^2}{m_j^2} - \frac{1}{(p-k)^2-m_\phi^2}\left( (2p-k)^2-\frac{1}{m_j^2}\left(2p\cdot k-k^2\right)^2\right) \right] \Bigg\}
\end{split}
\end{align}
On-shell, $p^2=m_\phi^2$, this expression reduces appreciably ($c_0\equiv 1, m_0 \equiv 0$):
\begin{align}
-i \, M^2(p^2=m_\phi^2) &= g^2 \sum\limits_{j=0}^{N-1} \int\frac{\dd^d k}{(2\pi)^d} \frac{c_j}{k^2-m_j^2} \left[ d - \frac{(2p-k)^2}{k^2-2p\cdot k} \right] \, .
\end{align}
Combining denominators and moving to Euclidean space, one can perform the loop integration via dimensional regularization: 
\begin{align}
-iM^2(p^2=m_\phi^2) = -ig^2 \sum\limits_{j=0}^{N-1} c_j \int\frac{\dd^d k_E}{(2\pi)^d}\bigg( \frac{d}{k_E^2+m_j^2} + \int_0^1\dd x  \frac{(x-2)^2m_\phi^2 - k_E^2 }{(k_E^2 + \Delta)^2} \bigg) \nonumber \\
= -ig^2 \sum\limits_{j=0}^{N-1} c_j \bigg[ -\frac{m_j^2}{2\pi^2} \times \frac{1}{\epsilon} + \text{finite} + \int_0^1\dd x  \bigg( \frac{2\Delta + (x-2)^2 m_\phi^2}{8\pi^2} \times \frac{1}{\epsilon} + \text{finite} \bigg) \bigg]
\end{align}
where we defined $\Delta = x^2m_\phi^2 + (1-x) m_j^2$.  The divergences, proportional to $m_\phi^2$ and $m_j^2$, cancel as per Eq.~\eqref{eq:cancellation}, similar to the higher-derivative calculation. The finite terms that do not vanish for the same reason are given by
\begin{align}
-iM^2(p^2=m_\phi^2) &= -\frac{ig^2}{(4\pi)^2} \sum\limits_{j=0}^{N-1} c_j \Bigg\{ 4 m_j^2\log m_j^2 - \int\limits_0^1\dd x \left[ 2\Delta+(x-2)^2m_\phi^2 \right] \log\Delta \Bigg\} \, .
\end{align}
The Feynman parameter integral can be evaluated in closed form and the final result is 
\begin{align}
\begin{split}
\label{eq:msquared-lw}
M^2(p^2=m_\phi^2) &= \frac{g^2}{(4\pi)^2} \sum\limits_{j=1}^{N-1} c_j \left\{ \frac12 \frac{m_j^4}{m_\phi^2} \log m_j^2 + \frac12 \frac{m_j \mu_j^3}{m_\phi^2} \log\left[ \frac{(m_j-\mu_j)^2}{4m_\phi^2} \right] \right\} \, , \\
c_j &\equiv (-1) \prod\limits_{\substack{k=1\\k\not=j}}^{N-1} \frac{m_k^2}{m_k^2-m_j^2} \, , \quad \mu_j \equiv \sqrt{m_j^2-4 \,m_\phi^2} \, ,
\end{split}
\end{align}
in agreement with Eq.~\eqref{eq:msquared-hd}. This provides a nontrivial cross check given the different choice of fixing the gauge after the field redefinitions that lead to the Lee-Wick basis.

\subsection{Explicit parametrization}
\begin{figure}[!htb]
\centering
\includegraphics[width=0.7\textwidth]{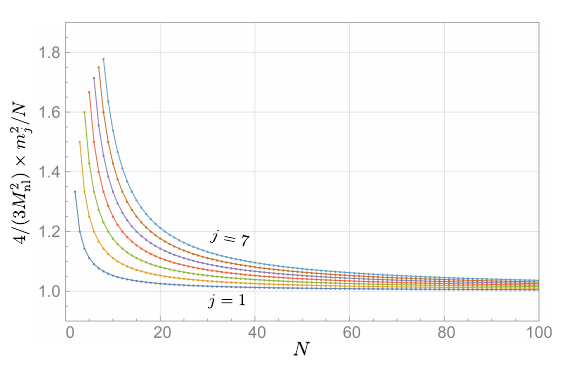}
\caption{Seven lightest masses of the explicit mass parametrization \eqref{eq:mass-parametrization}, normalized to the asymptotically nonlocal scale $M_\text{nl}$. For larger $N$ they all approach one common value, as required.}
\label{fig:mass-parametrization}
\end{figure}

In the scalar model of Ref.~\cite{Boos:2021chb}, the emergent nonlocal scale that regulates loop diagrams becomes relevant when $N \rightarrow \infty$ and the nondegenerate $m_j^2/(N-1)$ approach a common value.  Here we study Eq.~\eqref{eq:msquared-lw} using the same parametrization of the mass spectrum as in our scalar theory (see Fig.~\ref{fig:mass-parametrization} for a visualization),
\begin{align}
\label{eq:mass-parametrization}
m_j^2 = \frac{3}{2} \frac{N M_\text{nl}^2}{2-\frac{j}{N}} \, .
\end{align}
We are interested in the limit of $N\rightarrow\infty$, that is, $m_j\rightarrow\infty$, so it is useful to introduce the dimensionless quantity $\eta_j \equiv m_\phi/m_j$. Then the self-energy takes the form
\begin{align}
-iM^2(m_\phi^2) = \frac{ig^2 m_\phi^2}{(4\pi)^2} \sum\limits_{j=1}^{N-1} c_j \frac{1}{\eta_j^4} \left[ \log \eta_j - (1-4\eta_j^2)^{3/2} \log\left( \frac{1-\sqrt{1-4\eta_j^2}}{2\eta_j} \, \right) \right] \, .
\end{align}
For small arguments, $\eta_j \rightarrow 0$, again making use of the cancellation rules \eqref{eq:cancellation}, one has
\begin{align}
-iM^2(m_\phi^2) &\approx \frac{6ig^2 m_\phi^2}{(4\pi)^2} \sum\limits_{j=1}^{N-1} c_j \left[ \left( \frac{1}{\eta_j^2} - 1 \right) \log\eta_j + \frac34 \right] \, .
\end{align}
Re-inserting the masses $m_j$ and utilizing the parametrization Eq.~\eqref{eq:mass-parametrization} we can show that
\begin{align}
\lim\limits_{N\rightarrow\infty} \left[ -iM^2(m_\phi^2) \right]  &= -\frac{3ig^2}{(4\pi)^2}\lim\limits_{N\rightarrow\infty} \left( \frac32 m_\phi^2 + \sum\limits_{j=1}^{N-1} c_j m_j^2 \log\frac{m_j^2}{m_\phi^2} - m_\phi^2 \sum\limits_{j=1}^{N-1} c_j \log\frac{m_j^2}{m_\phi^2} \right) \\
&= -\frac{3ig^2}{(4\pi)^2} \left[ M_\text{nl}^2 + m_\phi^2 \left( \log \frac{M_\text{nl}^2}{m_\phi^2} + \frac32 - \gamma \right) \right] \, , \label{eq:self-energy-asnl}
\end{align}
where the last equality follows from the relations 
\begin{align}
\label{eq:numerical-formula-1}
&\sum\limits_{j=1}^{N-1} c_j m_j^2 \log \frac{m_j^2}{m_\phi^2} \approx M_{\rm nl}^2 + \mathcal{O}\left(\frac{1}{N}\right) \, , \\
&\sum\limits_{j=1}^{N-1} c_j \log \frac{m_j^2}{m_\phi^2} \approx \gamma - \log \frac{M_\text{nl}^2}{m_\phi^2} + \mathcal{O}\left(\frac{1}{N}\right) \, ,
\end{align}
which were determined numerically. Here, $\gamma = 0.577216...$ is the Euler--Mascheroni constant. Note that the scale $m_\phi^2$ is arbitrary in Eq.~\eqref{eq:numerical-formula-1}, as a consequence of the cancellation rules, Eq.~\eqref{eq:cancellation}. Equation~\eqref{eq:self-energy-asnl} confirms that the correction scale is indeed set by the nonlocal scale $M_\text{nl}^2$, with a subleading logarithmic term. In the massless limit $m_\phi \rightarrow 0$ one recovers a finite result that, up to relabelling of couplings, essentially reproduces the exact analytical result obtained in Ref.~\cite{Boos:2021chb}.

\begin{figure}[!htb]
\centering
\includegraphics[width=0.7\textwidth]{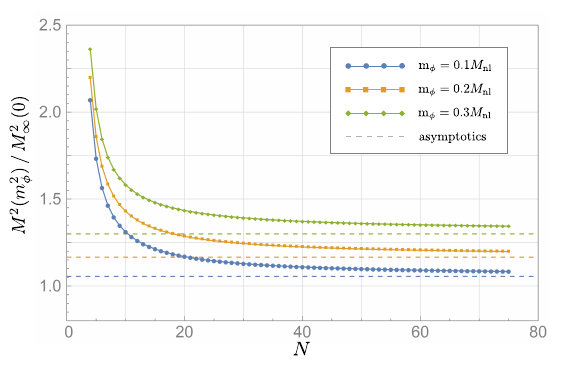}
\caption{Scalar self-energy at one loop for various scalar masses and $N$, normalized to $M_{\infty}^2(0) \equiv 3g^2M_\text{nl}^2/(4\pi)^2$. For larger $N$, the self-energy does not diverge; rather, it approaches a common, $m_\phi$-dependent value indicated by the dashed lines.}
\label{fig:self-energy}
\end{figure}

In order to verify \eqref{eq:self-energy-asnl}, we computed the self-energy numerically at finite $N$, for a range of scalar masses $m_\phi^2$, and 
compared this to the asymptotic value.  The results converge at large $N$, as shown in Fig.~\ref{fig:self-energy}.

\section{Asymptotically nonlocal non-Abelian theories and the Standard Model} \label{sec:sm}

In this section, we make some preliminary remarks on how our approach may be applied to more general gauge theories.  In Sec.~\ref{sec:scalarQED}, we considered a theory of a complex scalar field $\phi$ coupled to an 
asymptotically nonlocal Abelian gauge field.   At the one-loop level, it is not hard to see that we would obtain the same qualitative results for the scalar
self-energy if the gauge group in this theory were 
made non-Abelian; we comment on higher loops below.   For concreteness, we promote the U(1) gauge group to SU(N), and write
the non-Abelian field strength tensor using the matrix notation $F_{\mu\nu} \equiv T^a F^a_{\mu\nu}$.  Then, the gauge-invariant Lagrangian in higher-derivative form includes the terms
\begin{equation}
{\cal L} = -\frac{1}{2} \mbox{Tr } F_{\mu\nu}  \left[\prod_{j=1}^{N-1} \left(1+\frac{\ell_j^2 \Box}{N-1} \right) \right] F^{\mu\nu} - \phi^* \Box \phi + \cdots \,\, ,
\label{eq:nonabsc}
\end{equation}
where the $\Box$ operator is now built from covariant derivatives $\Box \equiv D_\mu D^\mu$, where $D_\mu = \partial_\mu - i  g \,T^a A^a_\mu$, and where the ellipsis includes the gauge fixing terms. 
Here, the scalar sector is local; we explain below how the result generalizes when the scalar sector is asymptotically nonlocal as well. To understand how the one-loop scalar self-energy computation is 
modified we note the following: (i)  the gauge boson propagator is determined by the quadratic terms in the Lagrangian, and hence has the same form as in the Abelian theory, aside from a Kronecker 
delta in the adjoint indices, and (ii) the vertices of the two diagrams that contribute to the scalar self-energy differ from the Abelian case only by factors of the group generators $T^a$, one for each gauge 
boson line emanating from a vertex.   Putting (i) and (ii) together, and using the fact that
\begin{equation}
{(T^a)_i}^j \,{(T^a)_j}^k = C_2(R) \, \delta_i^k  \,\,\, ,
\label{eq:casimir}
\end{equation}
we conclude that the only change to the one-loop results of Sec.~\ref{sec:scalarQED} is that there is now an overall multiplicative factor of the group invariant 
$C_2(R)$.  Note that $C_2(R) = \frac{1}{2 N} (N^2-1)$ for SU(N) if $R$ is the fundamental representation; the indices $i$ and $k$ in Eq.~(\ref{eq:casimir}) correspond to the ``colors" of 
the scalars on the external lines.   

The one-loop results also hold qualitatively if asymptotic nonlocality emerges in both the gauge and scalar sectors, {\em i.e.}, if we had
\begin{equation}
{\cal L} = -\frac{1}{2} \mbox{Tr } F_{\mu\nu}  \left[\prod_{j=1}^{N-1} \left(1+\frac{\ell_j^2 \Box}{N-1} \right) \right] F^{\mu\nu} - \phi^* \Box \left[\prod_{j=1}^{N-1} \left(1+\frac{\ell_j^2 \Box}{N-1} \right) \right]  \phi -m_\phi^2 \phi^*\phi +\cdots \,\, ,
\end{equation}
where we have assumed the same $\ell_j$ in both the scalar and gauge quadratic terms for simplicity, and the $\Box$ operator is again built from covariant derivatives.  The key point is that the scalar terms can
be rewritten in Lee-Wick form (just as in the $m_\phi=0$ example of Ref.~\cite{Boos:2021chb}), since our auxiliary field construction works
for any differential operator that is exponentiated in the nonlocal limiting theory.  Hence,
\begin{equation}
{\cal L} = -\frac{1}{2} \mbox{Tr } F_{\mu\nu}  \left[\prod_{j=1}^{N-1} \left(1+\frac{\ell_j^2 \Box}{N-1} \right) \right] F^{\mu\nu} + \sum_{j=0}^{N-1} (-1)^j \, \phi_j^* \, (\Box+m_j^2) \, \phi_j  \,\,\, ,
\label{eq:partlyLW}
\end{equation}
where the $m_j$ represent the scalar mass eigenvalues.  This leads to the same self-energy diagrams that we encountered previously for the local scalar field in Eq.~(\ref{eq:nonabsc}), but for each scalar mass 
eigenstate in Eq.~(\ref{eq:partlyLW}) separately, up to overall signs.

At higher-loop order, and in more general non-Abelian theories, the situation is more complicated. We argued in the context of the scalar theory of Ref.~\cite{Boos:2021chb} that our qualitative results on the 
separation between the emergent nonlocal scale and the lightest Lee-Wick mass should persist to any loop order, since there is only one scale in the limit of interest, $M_{\rm nl}$, that could serve as a regulator 
of the scalar self-energy.  This statement is supported by the explicit two-loop calculation that we include in the appendix.   More precisely, this dimensional argument is supported by the observation that both our 
asymptotically nonlocal $\phi^4$ and Abelian gauge theories are {\em finite} theories: in the formulation that we have presented, higher-derivative terms affect only propagators, so that these theories can be made 
arbitrarily more convergent than their local counterparts, and the nonlocal scale becomes the sole regulator in the theory.  The situation is less obvious in non-Abelian theories since gauge invariance 
requires higher-derivative interaction terms when one modifies the quadratic terms in the theory.  As noted in the Lee-Wick Standard Model~\cite{Grinstein:2007mp}, and illustrated in the $N=3$ model of 
Ref.~\cite{Carone:2008iw}, the higher-derivative interactions lead to (at most) logarithmic divergences, no matter how many additional derivatives are added via the quadratic terms.  Schematically, one might 
expect the on-shell scalar self-energy to have the form
\begin{equation}
-i M^2 (m_\phi^2) =  F_1(\{m_j\},m_\phi, N) \log\Lambda +F_2(\{m_j\},m_\phi,N) + {\cal O}(1/\Lambda) \,\,\, ,
\end{equation}
where $\Lambda$ represents the scale of a dimensionful, gauge-invariant regulator (like Pauli-Villars), and the $F_i$ are functions of the physical masses in the theory.  We may reasonably conjecture that 
\begin{equation}
F_i(\{m_j\},m_\phi, N) \rightarrow F_i(M_{\rm nl},m_\phi) \,\,\, 
\end{equation}
as $N \rightarrow \infty$, with the masses parametrized as in Eq.~(\ref{eq:mass-parametrization}).   In other words, functions of the physical masses should become functions of
$M_{\rm nl}$ and $m_\phi$ since no other dimensionful physical parameters appear in the Lagrangian as the limiting theory is approached.    This would be sufficient to address the hierarchy problem as $N$
becomes large and the Lee-Wick states are decoupled, since the dependence on the high scale $\Lambda$ remains logarithmic.  Diagrammatic verification of this conjecture in a dedicated study of asymptotically nonlocal non-Abelian theories will be deferred to future work.   

A separate challenge in the non-Abelian case is identifying an intermediate auxiliary field Lagrangian of a form similar to Eq.~\eqref{eq:aux-lagrangian} that connects the 
higher-derivative and Lee-Wick forms of a non-Abelian gauge sector for arbitrary $N$.  Finding a gauge-invariant Lagrangian of this form is no longer obvious, as Eq~\eqref{eq:abelian-gauge-trafo} is replaced by the more 
complicated non-Abelian field transformation.    For a specific choice of $N$, an auxiliary field representation of the non-Abelian gauge sector can be found with some effort, as was shown in the $N=3$ 
example in Ref.~\cite{Carone:2008iw}; we do not know whether it is possible to find a simple form for arbitrary $N$.  We also leave this problem for future work.  In the meantime, 
the higher-derivative form of the non-Abelian gauge kinetic terms is sufficient for constructing asymptotically nonlocal theories suitable for phenomenological study.   For example, one may write an 
asymptotically nonlocal version of the standard model as follows:
\begin{equation}
{\cal L} = {\cal L}_{\rm kin} + {\cal L}_{\rm Yuk} -V(H)  \,\,\, ,
\label{eq:sm1}
\end{equation}
where $ {\cal L}_{\rm Yuk}$ and $V(H)$ represent the usual standard model Yukawa couplings and Higgs doublet potential, respectively, while
the kinetic terms ${\cal L}_{\rm kin}$ are modified:
\begin{align}
{\cal L}_{\rm kin} & = - H^\dagger \Box f(\Box) H - \frac{1}{2} \mbox{Tr }G_{\mu\nu} f(\Box) G^{\mu\nu}- \frac{1}{2} \mbox{Tr }W_{\mu\nu} f(\Box) W^{\mu\nu}
- \frac{1}{4} B_{\mu\nu} f(\Box) B^{\mu\nu}  \nonumber \\ & + \sum_f \overline{f}_L i \slashed{D} f(\Box) f_L + \overline{f}_R i \slashed{D} f(\Box) f_R  \,\,\, ,
\label{eq:sm2}
\end{align} 
where
\begin{equation}
 f(\Box)=\prod_{j=1}^{N-1} \left(1+\frac{\ell_j^2 \Box}{N-1} \right) \,\,\, .
\label{eq:fofbox}
\end{equation}
Here $G_{\mu\nu}$, $W_{\mu\nu}$ and $B_{\mu\nu}$ represent the field strength tensors for the SU(3)$_C$, SU(2)$_W$ and U(1)$_Y$ gauge groups,
respectively, and the $\Box$ operator is built from the standard model covariant derivative, for example
\begin{equation}
D_\mu = \partial_\mu -i g_3 T^A g^A_\mu -i g_2 \frac{\sigma^a}{2} W^a_\mu -i g_Y B_\mu \,\,\, 
\end{equation}
for a matter field that is charged under all three gauge group factors. The sum over $f$ ranges over the set of standard model fermion 
fields.\footnote{Note that in principle there could be different asymptotically nonlocal scales for each kinetic term appearing in the Lagrangian.  The same is true for the possible Lee-Wick mass scales in the LWSM, where a common one is chosen for simplicity.}

It is worth noting that one can easily construct an auxiliary field formulation for an asymptotically nonlocal fermion sector.   Consider the following Lagrangian for $N$ left-handed
fermions $\psi^j_L$ and $N-1$ auxiliary right-handed fermions $\chi^j_R$:
\begin{equation}
\mathcal{L}_N = i \, \overline{\psi}^1_L \slashed{D} \,\psi^N_L - V(\psi^1_L) - \left\{ \sum_{j=1}^{N-1} \overline{\chi}^j_R \, \left[ \Box \psi^j_L - (\psi^{j+1}_L-\psi^j_L)/a_j^2\right] + \mbox{ H.c.} \right\}\,\,\, ,
\end{equation}
Integrating out the $\overline\chi^j_R$ gives the $N-1$ relations
\begin{align}
\psi^{j+1}_L = (1 + a_j^2 \Box) \, \psi^j_L \, .
\end{align}
Defining $\ell_j \equiv (N-1) \, a_j$, one has
\begin{align}
\psi^N_L = \prod\limits_{j=1}^{N-1}\left(1 + \frac{\ell_j \Box}{N-1} \right) \psi^1_L  \, .
\end{align}
A similar construction can be applied to fields with the opposite chirality, yielding fermionic terms of the form shown in Eq.~\eqref{eq:sm2}.  As we have seen before, $f(\Box) \rightarrow e^{\ell^2\Box}$, 
as $N \rightarrow \infty$, provided the $\ell_j \rightarrow  \ell$ in the same limit.  The construction summarized above also makes it straightforward to apply our previous asymptotically nonlocal generalization of scalar QED to QED itself.

It is interesting to note that the asymptotically nonlocal standard model Lagrangian given in Eqs.~(\ref{eq:sm1}) and (\ref{eq:sm2}) has the property that tree-level scattering amplitudes will have a momentum 
dependence that begins to deviate from standard model expectations in a way indistinguishable from the nonlocal limiting theory, when $N$ is large. This feature may lead to observable consequences at collider
experiments.

\section{Conclusions} \label{sec:conc}

In this paper, we have shown how to construct gauge theories that exhibit asymptotic nonlocality, extending previous work~\cite{Boos:2021chb} that was limited to theories of real scalar fields 
with $\phi^4$ interactions.  Asymptotically nonlocal theories represent a sequence of higher-derivative theories that approach a ghost-free nonlocal theory as a limit point.  Since the theories 
in this sequence involve finite numbers of derivatives, they avoid some of the technical complications inherent to infinite derivative theories, but nonetheless exhibit some of their distinctive features.   For example, in the scalar 
theories previously studied, loop diagrams are regulated by an emergent scale that does not appear as a fundamental parameter in the Lagrangian, corresponding to the nonlocal scale 
$M_{\rm nl}$ that is defined in the limiting theory.  As the limit is approached, Lee-Wick resonances become more plentiful in number but also decouple; one finds the relation 
\begin{align}
M_\text{nl}^2 \sim {\cal O}\left(\frac{m_1^2}{N}\right) \, ,
\end{align}
where $m_1$ is the lightest Lee-Wick resonance, and $N$ is the total number of poles in the two-point function.    In the $\phi^4$ theory that we previously studied~\cite{Boos:2021chb}, this 
parametric suppression implies that the scale of quadratic divergences may be held fixed as the Lee-Wick particles are taken heavy, something not possible in theories with a minimal spectrum of Lee-Wick particles.  Precisely the same behavior was found in the asymptotically nonlocal Abelian gauge theory that we studied in the present work.    

In particular, we studied the on-shell one-loop self-energy of a complex scalar explicitly in both the higher-derivative and Lee-Wick forms of asymptotically nonlocal scalar QED, where the latter 
formulation involves distinct fields for each physical particle, but no higher-derivative terms.   Like the purely scalar theory that we studied previously, we argued that our qualitative conclusions should 
hold to arbitrary loop order on dimensional grounds,  as $M_{\rm nl}$ is the only scale available that can regulate loop diagrams in the limiting theory, which is a finite quantum field theory.   As a 
nontrivial check, we supported this claim via an explicit two-loop calculation in asymptotically nonlocal $\phi^4$ theory, presented in the appendix of this paper.  We then showed how asymptotically 
nonlocal non-Abelian theories could be defined in higher-derivative form, assuring the existence of an emergent nonlocal scale as the Lagrangian approaches its limiting form; we presented the 
corresponding generalization of the standard model Lagrangian as a point of reference for future investigation.   Although asymptotically non-Abelian theories are not finite field theories (due to 
derivative interactions), we presented a plausibility argument for why the hierarchy problem may also be solved in these theories when the Lee-Wick particles are heavy, as we anticipate the 
scale of scalar self energies to be set by $M_{\rm nl}$ with at most logarithmic dependence on any higher cutoff.  We defer to future work a diagrammatic evaluation of this conjecture, and the 
related algebraic challenge of finding an auxiliary field description of asymptotically nonlocal non-Abelian theories that is valid for arbitrary $N$.

Whether this class of theories we study here can be made fully realistic as an extension of the standard model that addresses the hierarchy problem, while pushing some or all of the new
resonances outside the prying eyes of the LHC, remains an open question.   We hope the present work has laid the groundwork to consider this and related technical issues in future work.

\begin{acknowledgments}  
We thank the NSF for support under Grants PHY-1819575 and PHY-2112460.
\end{acknowledgments}

\appendix

\section{Asymptotically nonlocal $\phi^4$ theory at two loops} \label{app:2-loop}
In our review of asymptotically nonlocal $\phi^4$ theory in Sec.~\ref{sec:review}, we noted that the separation between the emergent nonlocal scale and the lightest 
Lee-Wick partner particle should persist at any order in perturbation theory, since the asymptotic form of the Lagrangian provides only one dimensional scale that could serve 
as a regulator of loop diagrams.   While this argument should be sufficient, we show as a consistency check that our conclusions remain unchanged if two-loop contributions to the mass renormalization are taken into account.\footnote{We thank the Referee of Ref.~\cite{Boos:2021chb} for challenging us to provide this example.} The three relevant diagrams are 
shown in Fig.~\ref{fig:sloops}, where diagram (a) was the one studied in Ref.~\cite{Boos:2021chb}.   We note that the sum of diagrams in (a) and (b) are nothing more than diagram (a) with the scalar propagator replaced by a ``dressed'' propagator:
\begin{figure}
      \centering
      \begin{subfigure}[b]{0.3\textwidth}
         \includegraphics{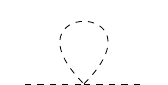}
         \caption{}
         \label{fig:1}
      \end{subfigure}
      \hfill
      \begin{subfigure}[b]{0.3\textwidth}
         \includegraphics{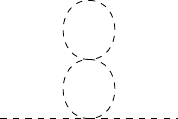}
         \caption{}
         \label{fig:2}
      \end{subfigure}
      \hfill
      \begin{subfigure}[b]{0.3\textwidth}
         \includegraphics{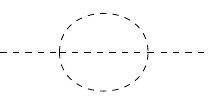}
         \caption{}
         \label{fig:3}
      \end{subfigure}
         \caption{Diagrams that contribute to the scalar mass renormalization through two loops.}
         \label{fig:sloops}
\end{figure}
\begin{equation}
D_F(p^2) = \frac{i}{p^2 \,  \prod_{j=1}^{N-1} \left(1-\frac{\ell_j^2 \, 
p^2}{N-1}\right)} \longrightarrow \frac{i}{p^2 \,  \prod_{j=1}^{N-1} \left(1-\frac{\ell_j^2 \, 
p^2}{N-1}\right) - M^2(p^2)} \,\, ,
\end{equation}
where $M^2(p^2)$ is the one-loop scalar self-energy, an $N$-dependent constant that we will call $m_0^2(N)$ below, which approaches the value given in 
Eq.~(\ref{eq:nlcutoff}) as $N \rightarrow \infty$, namely $m_0^2(\infty) = 3 \lambda \, M_{\rm nl}^2 / (128 \pi^2)$.   To determine the asymptotic value of diagrams (a) and (b),
we may again use the identity quoted in Ref.~\cite{Boos:2021chb},
\begin{align}
\lim\limits_{N\rightarrow\infty} \prod\limits_{j=1}^{N-1} \left( 1 + a_j^2 p_E^2 \right) = e^{\ell^2 p_E^2} \left[ \, 1 - \left(1+\frac{14}{27} \, p_E^2 \ell^2\right)
\frac{p_E^2 \ell^2}{N} + \mathcal{O}\left(\frac{1}{N^2}\right) \, \right] 
\, ,
\label{eq:expf}
\end{align}  
to obtain the generalization of Eq. (4.15) appearing in the same reference:
\begin{align}
\label{eq:aplusb}
\lim\limits_{N\rightarrow\infty} M^2(k^2)_{a+b} = \frac{\lambda}{16\pi^2} \int_0^\infty  \dd p_E \, p^3_E \, \frac{1}{p_E^2\,e^{\ell^2 p_E^2}+m_0^2(\infty)} \,\,\,.
\end{align}
Notice that this reduces to the asymptotic form for the one-loop self-energy if $m_0^2(\infty)$ is set to zero.  To understand the effect of the two-loop contribution we note that
\begin{equation}
\lim\limits_{N\rightarrow\infty} M^2(k^2)_{a+b} / M^2(k^2)_a = 2 \int_0^\infty \dd x \, \frac{x^3}{x^2 e^{x^2} + \frac{\lambda}{32 \pi^2}} \,\,\ ,
\end{equation}
where we have defined a dimensionless integration variable $x \equiv \ell \, p_E$.  The function on the right-hand side has an upper bound of $1$, which implies that the two-loop correction from 
diagram (b) does not change conclusion of Ref.~\cite{Boos:2021chb}, that the scale of the result is set by the nonlocal scale $M_{\rm nl}^2$.

The remaining diagram (c) is less trivial.  To consider its contribution to the scalar mass, we set the external momentum to zero, equivalent to evaluating the self-energy on-shell; in this case, expressing the propagators in the Lee-Wick basis (or equivalently, using a partial fraction decomposition of the propagators in the higher-derivative theory), the contribution to the amplitude may be written as
\begin{equation}
-i M^2(0)_c = \frac{i\, \lambda^2}{3!} \sum_{i,j,k=0}^{N-1} c_i c_j c_k \, I_{ijk}  \,\,\, ,
\label{eq:sumsumsum}
\end{equation}
where
\begin{equation}
I_{ijk} = \int \frac{\dd^d \ell}{(2\pi)^d} \int \frac{\dd^d p}{(2 \pi)^d} \, \frac{1}{p^2-m_i^2}\, \frac{1}{(p+\ell)^2-m_j^2} \,  \frac{1}{\ell^2-m_k^2} \,\,\, .
\label{eq:bigI}
\end{equation}
Here, $m_0=0$, corresponding to the massless theory studied in Ref.~\cite{Boos:2021chb}; the $c_i$ are defined in Eq.~(\ref{eq:cj}) and satisfy
the same cancellation rules (\ref{eq:cancellation}).

The ultraviolet divergences of Eq.~(\ref{eq:bigI}) will cancel in Eq.~(\ref{eq:sumsumsum}), as a consequence of Eq.~(\ref{eq:cancellation}), leaving a finite result.  Nevertheless,
it is useful if these divergences can be isolated cleanly at an intermediate step;  evaluation of Eq.~(\ref{eq:bigI}) by standard methods does not
provide for such a simple separation, as the divergences live partly in divergent Feynman parameter integrals.  A more tractable final form
can be obtained by rewriting Eq.~(\ref{eq:bigI}) using a trick~\cite{Kleinert:2001ax}:  one inserts the identity
\begin{equation}
1 = \frac{1}{2 \, d} \left(\frac{\partial \ell^\mu}{\partial \ell^\mu} + \frac{\partial p^\mu}{\partial p^\mu} \right) \,\,\, ,
\end{equation}
and then integrates the two terms by parts.   The result can be manipulated algebraically to show that the original integral may be reexpressed
as 
\begin{equation}
I_{ijk} =  \frac{3 \, m_j^2}{d-3} \, \int \frac{\dd^d \ell}{(2\pi)^d} \int \frac{\dd^d p}{(2 \pi)^d}  \frac{1}{
(p^2-m_i^2) [(p+\ell)^2-m_j^2]^2(\ell^2-m_k^2)} \,\,\, ,
\end{equation}
where we have used the freedom to relabel indices $i$, $j$ and $k$ using the total symmetry of $c_i \, c_j \, c_k$.  This
integral evaluates to
\begin{equation}
I_{ijk} = - \frac{3 m_j^2}{d-3} \frac{\Gamma(4-d)}{(4 \pi)^d}  \int_0^1 \dd x \, [x(1-x)]^{\frac{d}{2}-2}
\int_0^1 \dd w \, w (1-w)^{1-\frac{d}{2}} \, \Delta^{d-4}  \,\,\, ,
\label{eq:bigId}
\end{equation}
where 
\begin{equation}
\Delta = w \, m_j^2 + \frac{(1-w)}{x(1-x)}[x \, m_i^2 + (1-x) m_k^2] \, .
\end{equation}
One may now expand  with $d=4-\epsilon$.  Writing the result in terms of its divergent and finite parts, we find
\begin{equation}
I_{\rm div} = \frac{3 m_j^2}{(4 \pi)^4} \left[ \frac{2}{\epsilon^2} +\frac{1}{\epsilon} \left( 3 -2 \gamma + 2 \ln(4 \pi / m_j^2) \right)
\right] \,\,\, .
\end{equation}
Since this expression is entirely independent of the indices $i$ and $k$,
\begin{equation}
\sum_{i,k=0}^{N-1} c_i \, c_k \, I_{\rm div} = 0  \,\,\, ,
\label{eq:vansum}
\end{equation}
and there are no divergent contributions to the amplitude in Eq.~(\ref{eq:sumsumsum}). The finite part of Eq.~(\ref{eq:bigId}) is given by
\begin{equation}
I_{\rm finite} = - \frac{3 \, m_j^2}{(4 \pi)^4} \int_0^1 \dd x \int_0^1 \dd w \, \frac{w}{1-w} \ln(\Delta/m_j^2) + \cdots \,\,\, ,
\end{equation}
where the ellipsis represents terms that vanish under the same summation Eq.~(\ref{eq:vansum}). Discarding those terms we are led to our final result
\begin{align}
\begin{split}
& -i M^2(0)_c = \frac{-i\, \lambda^2}{2(4\pi)^4} \sum_{i,k=0}^{N-1} \sum_{j=1}^{N-1} c_i c_j c_k \, m_j^2 \tilde{I}_{ijk} \, , \\
& \tilde{I}_{ijk} = \int_0^1 \dd x \int_0^1 \dd w \frac{w}{1-w} \log\left[ w + (1-w)\frac{x \, m_i^2 + (1-x) m_k^2}{x(1-x) m_j^2} \right] \, .
\end{split}
\end{align}
Assuming the mass parametrization given in Eq.~(\ref{eq:solvable-model-mass-parametrization}), the convergent integral $\tilde{I}_{ijk}$ can be evaluated numerically, and one finds that it is roughly of order unity.  However,
one also finds that the triple sum involves significant cancellations between large terms of alternating sign, as a result of the properties of the coefficients $c_i$, $c_j$, and $c_k$.  As a result, the
integral $\tilde{I}_{ijk}$ must be evaluated to high precision in order to obtain accurate results; convergence is slow and worsens as $N$ becomes large.    Nevertheless,  one can see the expected results emerging in numerical data set shown in Table~\ref{table:c} below, where we have chosen some points between $N=5$ and $20$ for illustration.  The results are given in units of 
$\lambda^2 M_{\rm nl}^2/(512\pi^4)$.
\begin{table}[h!]
\centering
\begin{tabular}{lclclclclclcl} \hline \hline
$N$        && 5      && 6      && 8      && 10     && 15     && 20     \\ \hline
$M^2(0)_c$ && 0.4019 && 0.3771 && 0.3501 && 0.3356 && 0.3182 && 0.3099 \\ \hline \hline
\end{tabular}
\caption{Contribution of diagram (c) to the self-energy, normalized by $\lambda^2 M_{\rm nl}^2/(512\pi^4)$.} 
\label{table:c}
\end{table}

Notice that in the case where $N=20$ the lightest Lee-Wick partner squared mass is four times as large compared to the case where $N=5$. The value of diagram (c), however, shows no corresponding growth and remains of the same order. This indicates that quadratic divergences are not reemerging at the two-loop level, supporting the dimensional argument that they will not do so at any order in the loop expansion for this theory.

\end{document}